\documentclass[aps,pra,twocolumn,superscriptaddress]{revtex4-2}
\usepackage{graphicx}
\usepackage{amsfonts}
\usepackage{amsmath}

\usepackage[usenames,dvipsnames]{color}
\usepackage{bm}
\usepackage{amssymb}
\usepackage{braket}

\usepackage{multirow}
\usepackage{epstopdf}
\usepackage[normalem]{ulem}

\usepackage{hyperref}
\hypersetup{
colorlinks=true,
linkcolor=blue,
citecolor=blue,
filecolor=green,
urlcolor=blue,
}
\usepackage[nameinlink,capitalise]{cleveref}
\usepackage[caption=false]{subfig}

\begin{document}


\title{Magnetic Feshbach resonances in ultracold atom-molecule collisions}


\author{Masato Morita}
\affiliation{Chemical Physics Theory Group, Department of Chemistry, and Center for Quantum Information and Quantum Control, University of Toronto, Toronto, Ontario, M5S 3H6, Canada}
\author{Maciej B. Kosicki}
\affiliation{Faculty of Physics, Kazimierz Wielki University, al. Powsta\'nc\'ow Wielkopolskich 2, 85-090 Bydgoszcz, Poland}
\author{Piotr S. {\.Z}uchowski}
\affiliation{Institute of Physics, Astronomy and Informatics, Nicolaus Copernicus University, Toru{\'n}, Poland,  87-100}
\author{Paul Brumer}
\affiliation{Chemical Physics Theory Group, Department of Chemistry, and Center for Quantum Information and Quantum Control, University of Toronto, Toronto, Ontario, M5S 3H6, Canada}
\author{Timur V. Tscherbul}
\affiliation{Department of Physics, University of Nevada, Reno, Nevada, 89557, USA}

\date{\today}

\begin{abstract}


We report numerically exact
quantum scattering calculations on magnetic Feshbach resonances in ultracold, strongly anisotropic atom-molecule [Rb($^2$S)~+~SrF($^2\Sigma^+$)] collisions based on state-of-the-art {\it ab initio} potential energy surfaces.
We find broad resonances mediated by the intermolecular spin-exchange interaction, 
 as well as narrow resonances due to the intramolecular spin-rotation interaction, which are unique to atom-molecule collisions.  Remarkably, the density of  resonances in atom-molecule collisions is not much higher than that in atomic  collisions despite the presence of a dense manifold of molecular rotational states, which can be rationalized by analyzing the adiabatic states of the collision complex.


 \end{abstract}
\maketitle
\newpage

{\it Introduction.} 
Magnetic Feshbach resonances (MFRs) are a powerful tool to control interatomic interactions in ultracold atomic gases \cite{Bloch:08,Chin:10} being at the very heart of many intriguing quantum phenomena, such as  the formation of 
weakly bound few-body states  \cite{Kohler:06,Greene:17},  quantum chaos in  ultracold  collisions of lanthanide atoms \cite{Frisch:14,Maier:15},  
and the BEC-BCS crossover \cite{Bloch:08,Gross:17}.
The physical mechanism responsible for MFRs relies on the coupling  between the open (i.e., scattering) and closed  (i.e., bound) states of the collision complex, which is greatly
amplified when these states become degenerate, leading to a resonance enhancement of the scattering cross section \cite{Chin:10}. This coupling can arise from a variety of mechanisms,  ranging from short-range spin-dependent  
and hyperfine interactions in ultracold atomic collisions \cite{Chin:10,Brue:12,Barbe:18,Frisch:14,Makrides:18}
 radiative transitions \cite{Fedichev:96,Bohn:97,Devolder:19}, dc electric fields \cite{Marinescu:98,Krems:06}, off-resonant
laser fields \cite{Crubellier:17}, and radiofrequency dressing \cite{Kaufman:09,Tscherbul:10b,Papoular:10}.
On the theoretical side, MFRs are most accurately described by rigorous coupled-channel (CC) quantum scattering calculations \cite{Chin:10}, which yield numerically exact solutions of the Schr\"odinger equation for a given interatomic interaction potential, thus providing unbiased insights into ultracold collision physics.

While MFRs in ultracold atom-atom collisions are well-understood \cite{Chin:10}, much less is known about their atom-molecule  \cite{Yang:19,Zhen:22,Son:22,Park:23b,Karman:23}  and  molecule-molecule \cite{Park:23} counterparts, which have recently been  observed by several groups. This is due to enormous computational challenges \cite{Morita:19b,Bause:23} caused by a large number of rotational, fine, and hyperfine states coupled by strongly anisotropic interactions, which are involved in ultracold collisions of trapped molecules produced by photoassociation (such as KRb and NaK \cite{Ni:08,Park:15,Bause:23,Bohn:17}) and laser cooling (such as SrF, CaF, and YO  \cite{Shuman:10,Barry:14,Truppe:17,Anderegg:18,McCarron:18,Ding:20}). 
Because the computational cost of solving CC equations scales steeply with the number of molecular rotational states ($N_\text{max}^6$ \cite{Truhlar:94}), their direct numerical solution  rapidly becomes computationally impractical \cite{Morita:19b,Bause:23} for all but the simplest atom-molecule collisions involving weakly interacting atoms (such as He, Mg, and N) \cite{Krems:03,Campbell:09,Hummon:11}.
 Restricted basis set computations have been performed on selected systems \cite{Wallis:14,Hermsmeier:21,Karman:23} yielding useful qualitative insights into MFRs in Na~+~NaLi collisions \cite{Hermsmeier:21,Karman:23}. However, numerically exact CC calculations on MFRs in ultracold atom-molecule collisions with realistic (strongly anisotropic) interactions have remained elusive due to the unresolved challenges mentioned above.
  At present, such calculations are limited to atom-molecule pairs, whose hyperfine structure can be neglected (e.g., by selecting fully spin-polarized initial states  \cite{Tscherbul:11,Morita:17,Morita:18}), precluding theoretical insights into a growing number of experimental observations of MFRs in ultracold atom-molecule collisions \cite{Yang:19,Zhen:22,Son:22,Park:23b} and the exploration of MFRs  in new molecular systems.


Here, we report the first numerically exact CC calculations of  MFR spectra in ultracold, strongly anisotropic atom-molecule collisions. We focus on ultracold Rb$(^2\mathrm{S})$~+~SrF($^2\Sigma^+$) collisions, which are experimentally relevant in several contexts. First, SrF($^2\Sigma^+$)  was the first molecule to be laser-cooled \cite{Shuman:10} and confined in a magneto-optical trap \cite{Barry:14}, and collisions with co-trapped Rb atoms have been proposed as a means to further cool SrF molecules by momentum-transfer collisions with ultracold alkali-metal atoms  \cite{Lim:15,Morita:18}, a technique known as sympathetic cooling \cite{Lara:06,Tscherbul:11,Lim:15,Morita:17,Morita:18,Zhang:24}.
 Second, ultracold collisions of closely related CaF$(^2\Sigma^+)$ molecules with co-trapped Rb atoms \cite{Jurgilas:21}, and  with other CaF molecules have already been observed \cite{Cheuk:20} and  suppressed \cite{Anderegg:21} using microwave shielding.  Our results could thus be verified in near-future ultracold collision experiments.
 
 The Rb-SrF collision complex is characterized by two deep and strongly anisotropic 
{potential energy surfaces (PESs)}
 of singlet and triplet symmetries  \cite{Kosicki:17},  which can couple hundreds of rotational states \cite{Morita:18}, in addition to the pronounced hyperfine structure of both SrF and Rb, causing major computational challenges, as mentioned above.
 Here, we address these challenges  using a recently developed total rotational angular momentum (TRAM) representation \cite{Tscherbul:23tram}, which enables us to  efficiently incorporate all rotational, fine,  and hyperfine states of the collision partners  in converged CC calculations on realistic {\it ab initio} PESs.
  Our calculated MFR spectra are dominated by broad resonances with a density of about 5 resonances per 1~kG, which should be readily observable experimentally. Our results open up the possibility of rigorous CC calculations of MFRs on a wide range of ultracold atom-molecule collisions of experimental interest. 
 

{\it Theory.} The Hamiltonian for a $^{2}\Sigma$ molecule (such as SrF) colliding with a $^2\mathrm{S}$ atom (such as Rb) may be written in  atomic units ($\hbar=1$) as \cite{Krems:04,Tscherbul:23tram} 
\begin{equation}
\hat{H}=-\frac{1}{2\mu R}\frac{\partial^2}{\partial R^2}R + \frac{\hat{L}^2}{2\mu R^2}
+ \hat{V}_\text{int} 
+ \hat{H}_\text{mol} + \hat{H}_\text{at},
\label{eq:H}
\end{equation}
where $\mu$ is the reduced mass of the atom-molecule trimer described by the Jacobi vectors $\mathbf{R}$ (pointing from the molecule's center of mass to the atom)  and $\mathbf{r}$  (joining the nuclei of the diatomic molecule), and $\hat{L}^2$ is the squared orbital angular momentum for the collision.
The atom-molecule interaction
{is given as}
$ \hat{V}_\text{int} = \sum_{S,M_S} V_{S}(R,\theta) |SM_S\rangle  \langle SM_S| + \hat{V}_\text{mdd} $, where  $V_S(R,\theta)$ 
{are} 
the singlet ($S=0$) and triplet ($S=1$) adiabatic PESs (see below), and  $\hat{V}_\text{mdd}$ 
{is}
the magnetic dipole-dipole interaction. We adopt the rigid-rotor approximation by setting $r=r_e$, the equilibrium distance in SrF,  which is expected to be qualitatively accurate for collisions of rigid molecules such as SrF \cite{Morita:20}.
 As such, the atom-molecule PES only depends on $R$ and $\theta$, the angle between $\mathbf{R}$  and $\mathbf{r}$.
In \cref{eq:H}, 
$\hat{H}_\text{mol}$ describes the isolated $^2\Sigma$ molecule, 
$\hat{H}_\text{mol}= \hat{H}_\text{mol}^\text{rot} + \hat{H}_\text{mol}^\text{spin} + \hat{H}_\text{mol}^\text{spin-rot}$,
where $\hat{H}_\text{mol}^\text{rot}  = B_e\hat{\mathbf{N}}^2$ is the rotational Hamiltonian, $B_e$ is the rotational constant, $\hat{\mathbf{N}}$ is the rotational angular momentum of the diatomic molecule, and $\hat{H}_\text{mol}^\text{spin}  = g_S\mu_0 B \hat{S}_Z +  a\hat{\mathbf{I}}\cdot\hat{\mathbf{S}}$ is the spin Hamiltonian, which accounts for the electron and nuclear spins within the molecule described by the operators $\hat{\mathbf{S}}$ and $\hat{\mathbf{I}}$ interacting with each other via the Fermi contact interaction (coupling constant $a$) and with an external magnetic field $B$ directed along a space-fixed $z$-axis. 
The spin-rotation interaction 
$\hat{H}_\text{mol}^\text{spin-rot}= \gamma_\text{sr} \hat{\mathbf{N}}\cdot \hat{\mathbf{S}} + \frac{c\sqrt{6}}{3}\bigl{(}\frac{4\pi}{5}\bigr{)}^{1/2}
  \sum_{q} (-1)^qY_{2-q}(\theta_r,\phi_r) [\hat{\mathbf{I}}\otimes \hat{\mathbf{S}}]^{(2)}_q$  includes the electron spin-rotation and anisotropic hyperfine interactions, parametrized by the coupling constants   $\gamma_\text{sr} $ and $c$. The internal structure of atomic Rb($^2$S) is described by the Hamiltonian $\hat{H}_\text{at}  = g_S\mu_0 B \hat{S_a}_{z} + A_a\hat{\mathbf{I}}_a\cdot\hat{\mathbf{S}}_a$, where ${\mathbf{S}}_a$ and $\hat{\mathbf{I}}_a$ are the atomic electron and nuclear spin operators, and $A_a$ is the hyperfine constant. 

To map out the spectrum of MFRs in ultracold Rb~+~SrF collisions, we solve the Schr\"odinger equation $\hat{H}|\Psi\rangle = E|\Psi\rangle$, where $E$ is the total energy and $|\Psi\rangle$ is the scattering wavefunction, by expanding the wavefunction of the atom-molecule collision complex as $|\Psi\rangle=R^{-1}\sum_i F_i(R) |\Phi_i\rangle$, where 
${|\Phi_i\rangle} =|(Nl)J_rM_r\rangle |n_s\rangle |n_s^{(a)}\rangle$ are the TRAM basis functions \cite{Simoni:06}, 
which enable a computationally efficient treatment of the essential hyperfine structure of the collision partners interacting via highly anisotropic potentials in an external magnetic field \cite{Tscherbul:23tram}.  Here, $|(Nl)J_rM_r\rangle$ are the eigenstates of  $\hat{J}_r^2$ and $\hat{J}_z$, where $\hat{\mathbf{J}}_r=\mathbf{N}+\mathbf{L}$ is the TRAM of the collision complex and $\hat{J}_z$ is its projection on the field axis. The atomic and molecular spin functions are denoted by  $|n_s^{(a)}\rangle$ and $ |n_s\rangle$. 


Unlike the total angular momentum of the collision pair, $J_r$ is rigorously conserved even in the presence of external magnetic fields and isotropic hyperfine interactions \cite{Tscherbul:23tram}. The different values of $J_r$ can only be coupled by weak spin-rotation interactions, such as the spin-rotation and anisotropic hyperfine interactions \cite{Tscherbul:23tram}. 
 The ensuing near-conservation of $J_r$ is key to the computational efficiency of the TRAM basis, which enables the inclusion of a sufficient number of closed rotational channels ($N_\text{max}=175$ \cite{SM}) to produce converged results for highly anisotropic Rb~+~SrF collisions.
We integrate the CC equations numerically  \cite{Johnson:73,Manolopoulos:86} 
to obtain the asymptotic behavior of the radial solutions $F_i(R)$, from which the $S$-matrix elements and scattering cross sections are obtained as a function of magnetic field \cite{SM,Tscherbul:23tram}.

{\it Ab initio calculations.} 
To describe the quantum dynamics of ultracold Rb~+~SrF collisions it is essential to have a realistic description of both the singlet and triplet PESs.
Here, we employ the state-of-the-art {\it ab initio} triplet PES \cite{Morita:18} computed using the coupled cluster method with a large basis set.
To determine the singlet PES, we employ the Complete Active Space Self-Consistent Field (CASSCF) method using an aug-cc-pVQZ basis set and ECP28MDF core potentials for Rb and Sr \cite{SM}. The active space for this calculation includes the open-shell orbitals of both Rb and SrF, accommodating two electrons in two orbitals. 
The exchange energy thus obtained is added to the triplet PES \cite{Morita:18} to produce the singlet PES.
 This procedure ensures the correct behavior of the singlet PES in the limit $R\to\infty$, where the exchange energy decays exponentially, and accurately accounts for dispersion interactions, which should asymptotically be identical for both the singlet and triplet PESs. We verified the correctness of the singlet-triplet energy splitting using additional spin-flip Symmetry Adapted Perturbation Theory (SAPT) calculations \cite{Patkowski:18}, in which the asymptotic behavior is calculated based on monomer properties \cite{SM}.

\begin{figure}[t!]
	\centering
	\includegraphics[width=1.0\columnwidth, trim = 0 0 0 0]{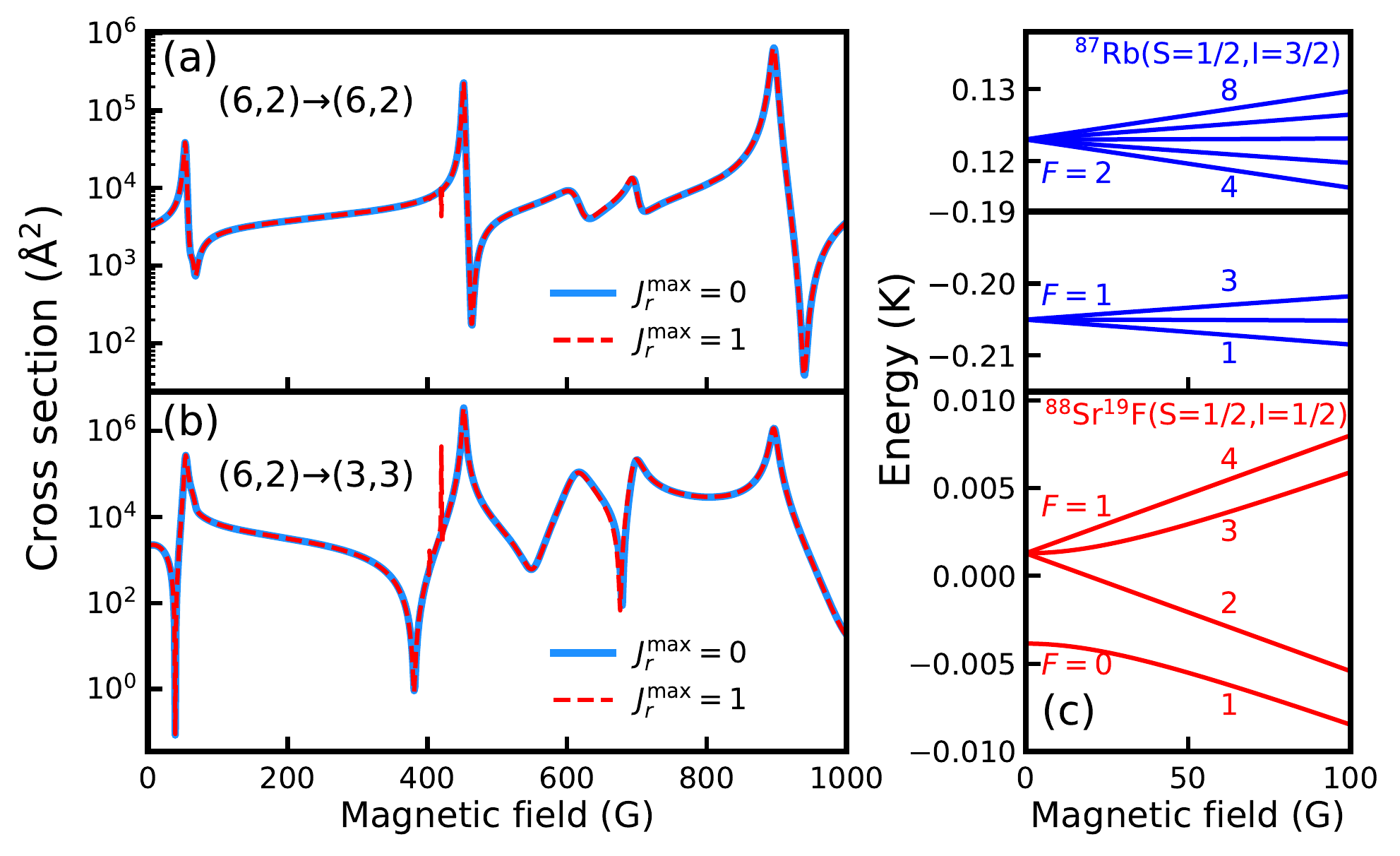}
	 \caption{ 
Magnetic field dependence of (a) elastic and (b) inelastic cross sections for Rb($|6\rangle$)~+~SrF($|2\rangle$) collisions calculated with 
$J_r^\text{max}=0$ (solid lines) and $J_r^\text{max}=1$  (dashed lines).
	 The collision energy is $E_\text{c}=0.1\, \mu$K. (c) Hyperfine-Zeeman energy levels of  Rb($^2$S) (upper panel) and SrF$(^2\Sigma^+)$ (lower panel) in its ground rovibrational state.}
\label{fig:ch}
\end{figure}

{\it Magnetic Feshbach resonances in ultracold atom-molecule collisions.} 
We consider ultracold Rb atoms and SrF molecules colliding in the initial states $\ket{i}\otimes \ket{j}$, where $\ket{i}$ and $\ket{j}$  stand for the initial hyperfine-Zeeman states of Rb and SrF,  labeled in the order of increasing energy, as shown in \cref{fig:ch}(c). In the low $B$ field limit, these states are identical to the atomic and molecular hyperfine states $\ket{F_a m_{F_a}}$ and $\ket{(NS I)F m_{F}}$, where $\hat{\mathbf{F}}=\hat{\mathbf{N}} + \hat{\mathbf{S}}+\hat{\mathbf{I}}$ is the total angular momentum of SrF, which is a vector sum of the rotational, electron spin, and nuclear spin angular momenta. We are interested in the experimentally relevant case of SrF molecules colliding in their ground rotational states $(N=0)$, which can be prepared by laser cooling and trapped in an optical dipole trap \cite{Barry:14,McCarron:18}. The hyperfine structure of SrF($N=0$) consists of two hyperfine manifolds with $F=1,0$, which are split by an external magnetic field into 4 levels [see \cref{fig:ch}(c)]. 
Non-fully spin-polarized initial states of Rb and SrF can undergo collision-induced spin-exchange, i.e., $\ket{6} \otimes \ket{2}$ $\to$ $\ket{3} \otimes \ket{3}$, which conserves the projection of the total internal angular momentum of the collision partners $m_{F_a}+m_{F}$.
 The rigorous treatment of the spin-exchange collisions and of the hyperfine structure of the collision partners constitute two key improvements over the previous calculations \cite{Morita:18}.

Figures~\ref{fig:ch}(a) and (b) show the integral cross sections for ultracold Rb~+~SrF collisions as a function of magnetic field calculated with two different basis sets containing the single lowest ($J_r^\text{max}=0$) and two lowest ($J_r^\text{max}=1$) TRAM blocks. Both of these basis sets are converged with respect to the maximum number of rotational basis states of SrF ($N_\text{max}=175$ \cite{SM}), so any difference between them quantifies the extent of TRAM conservation in ultracold Rb~+~SrF collisions.
We observe that, save for a single narrow feature near 400~G, the cross sections calculated using the two basis sets are essentially the same across the entire magnetic field range, confirming that $J_r$ is indeed a nearly good quantum number in Rb~+~SrF collisions. The broad MFRs occur due to the intermolecular spin-exchange interaction   \cite{SM} (see below), and the narrow peaks can be attributed to the coupling between the different values of $J_r$ induced by the spin-rotation interaction in SrF.
We will focus on the broad MFRs obtained from $J_r^\text{max}=0$ calculations.  



\begin{figure}[t!]
	\centering
	\includegraphics[width=0.9\columnwidth, trim = 0 20 0 -5]{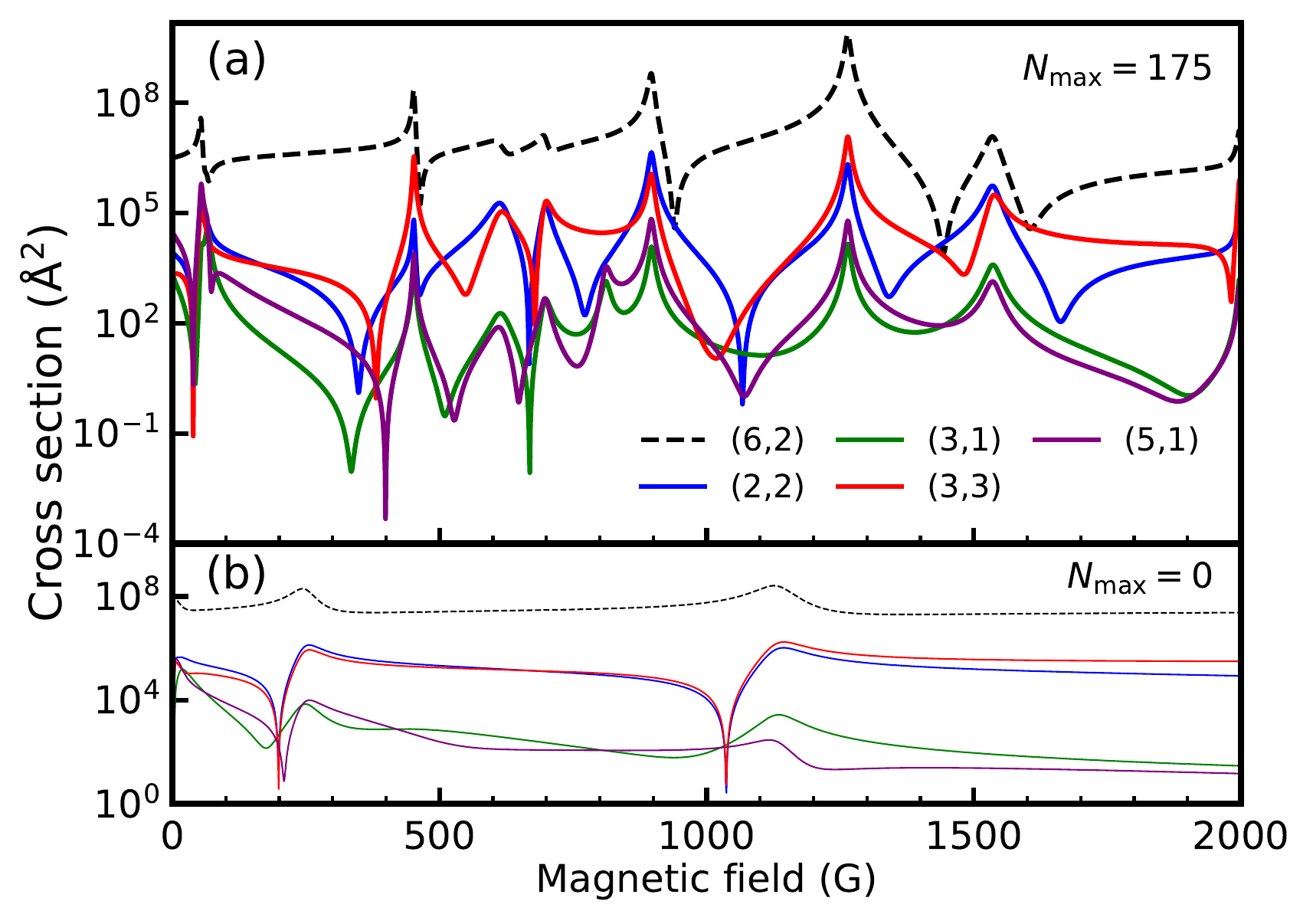}
	\caption{ 
	(a) Calculated MFR spectra  for elastic and inelastic Rb($|6\rangle$)~+~SrF($|2\rangle$) collisions at $E_\text{c}=0.1\, \mu$K.
The elastic cross section (black dashed line) is multiplied by 1000 for visual clarity. 
	The final states are marked as ($i,j$), where $i$ and $j$ label the internal states of Rb and SrF (see text and \cref{fig:ch}(c)).    
        (b) Same as  (a) but with only the $N=0$ rotational 
        {state}
        of SrF retained in the CC basis.}
\label{fig:state-to-state}
\end{figure}

In \cref{fig:state-to-state}(a), we show the calculated spectrum of MFRs in ultracold Rb~+~SrF collisions obtained from fully converged CC calculations with  1232 TRAM basis functions. 
{The inelastic and elastic cross sections are comparable, which indicates that 
 significant inelastic losses can occur near MFRs, making it hard to prepare a stable atom-molecule mixture. Further work is needed to identify the parameter regimes, where these losses are suppressed.}
The resonance density  (4-5 per 1~kG) is fairly insensitive to the  initial and 
final  hyperfine states of Rb and SrF, and is comparable to that of MFRs in ultracold Rb~+~Rb collisions \cite{Marte:02}.
The $J_r$-conserving MFRs are broad and similar to those which occur in ultracold collisions of alkali-metal atoms \cite{Chin:10}. This is consistent with a recent theoretical prediction \cite{Bird:23} of alkali-metal-like MFRs in ultracold Rb~+~CaF collisions based on model CC calculations, which neglected the atom-molecule interaction anisotropy. However, as shown below, the anisotropy has a non-negligible effect on MFR spectra.
The hyperfine structure of the collision partners and their spin-exchange interactions also play a crucial role: When the spin-exchange coupling  is omitted in test calculations, most of the MFRs shown in \cref{fig:state-to-state}(a) disappear \cite{SM}.

The relatively low density of MFRs in Rb~+~SrF collisions is surprising, given the very large number of rovibrational and hyperfine states of the Rb-SrF collision complex that need to be included in CC calculations to produce converged results.
\Cref{fig:state-to-state}(b) shows a 
reduction in the number of MFRs when the anisotropy of the Rb-SrF interaction if turned off by setting $N_\text{max}\to 0$, illustrating the significant role of the anisotropic coupling between the rotational states of SrF in enhancing the density of MFRs.
However, the enhancement is not dramatic. A similar observation has recently been made for MFRs in ultracold Na~+~NaLi collisions    \cite{Karman:23}. 

To further elucidate the role of anisotropic couplings between different rotational states, we diagonalized the atom-molecule Hamiltonian in \cref{eq:H} without the radial kinetic energy term to produce the adiabatic states of the Rb-SrF collision complex as a function of $R$.  
In the adiabatic picture, the Rb~+~SrF collision partners approach each other at long range on a smooth adiabatic potential, which experiences a series of avoided crossings with other potentials as  $R$ is decreased, leading to inelastic transitions and MFRs \cite{SM}.
The density of the adiabatic 
potentials in the vicinity of  
the incident collision threshold is  fairly well-converged already for moderate values of $N_\text{max}\simeq 30{-}60$. Further increasing $N_\text{max}$ only produces slight changes in the shapes of the existing adiabatic curves, 
 leading to small changes in closed-channel bound states, which are not enough to produce new bound states 
that could result in additional MFRs. Therefore, the MFR density saturates at $N_\text{max}\simeq 30{-}60$ \cite{SM}, well below the value required for full convergence ($N_\text{max}=175$). 
In other words,  {\it even though the adiabatic states themselves are not completely converged at low $N_\text{max}$,  their density near the incident threshold is}. This explains why highly excited rotational states do not contribute directly to the MFR density, although they do affect the positions and widths of the individual MFRs.

 \begin{figure}[t!]
	\centering
	\includegraphics[width=1.0\columnwidth, trim = 0 0 0 0]{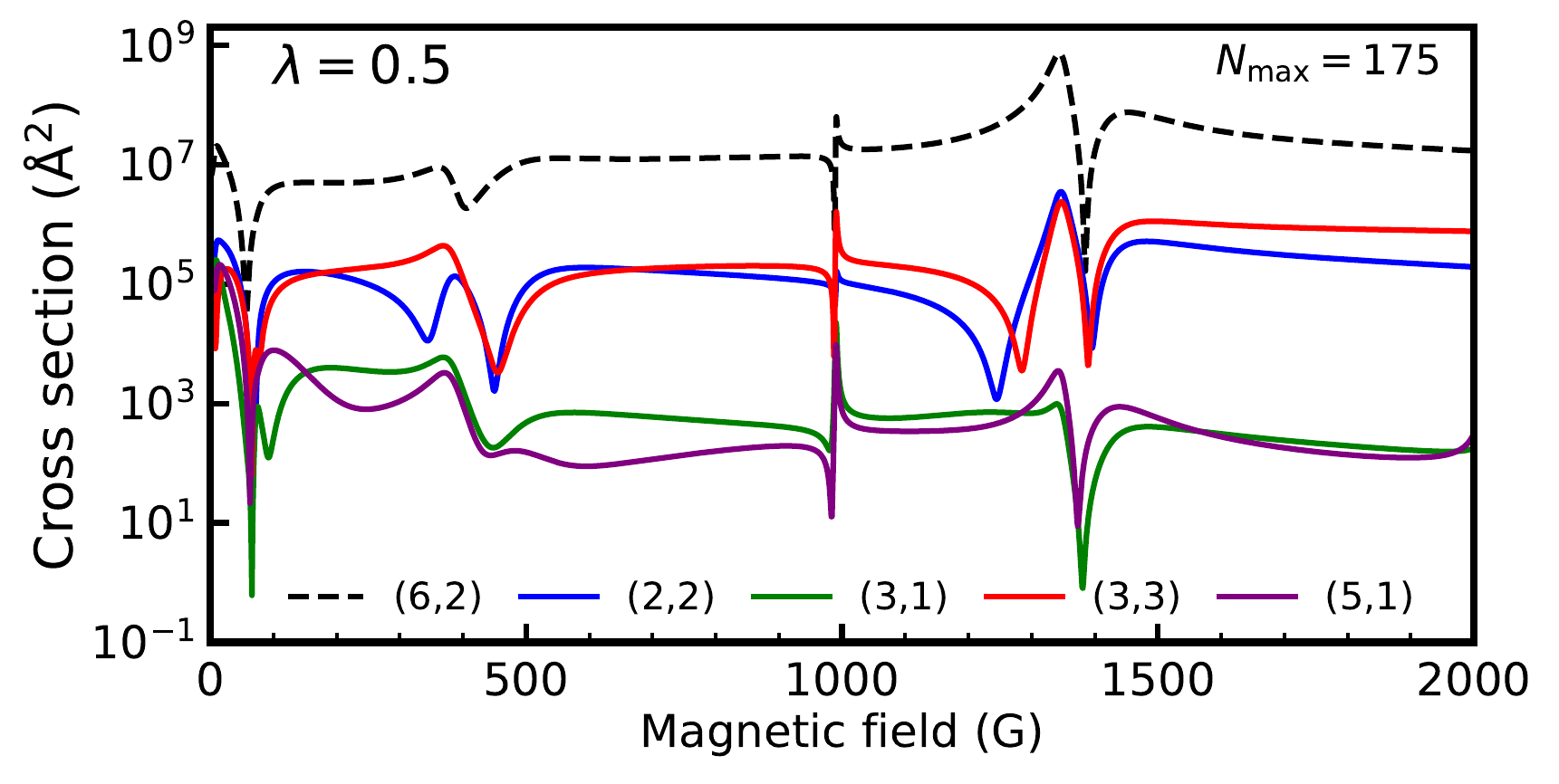}
	\caption{  
Calculated MFR spectra  for Rb($|6\rangle$)~+~SrF($|2\rangle$) collisions at $E_\text{c}=0.1\, \mu$K with the short-range interaction PES scaled by a factor of 0.5.	The elastic cross section (black dashed line) is multiplied by 1000 for visual clarity.}
\label{fig:scaled}
\end{figure}

It has been hypothesized that MFRs in ultracold collisions of alkali-dimers with alkali-metal atoms 
are caused by the long-range LeRoy-Bernstein bound states \cite{Frye:23}, which are only weakly affected by the atom-molecule interaction anisotropy, since it typically decreases much faster with $R$ than the isotropic part of the PES.
 If this hypothesis is true, the atom-molecule MFRs would be very similar to the well-known atom-atom MFRs  \cite{Chin:10} in that the effects of short-range interaction anisotropy can be safely neglected.  
An opposing hypothesis states that MFRs are due to the short-range  chaotic states of the atom-molecule collision complex \cite{Mayle:12,Mayle:13}.

{We emphasize that all calculations on atom-molecule MFRs reported thus far (see, e.g., \cite{Bird:23,Karman:23}) were performed with severely restricted rotational basis sets. They were thus affected by uncontrolled basis set truncation errors, which are absent in our calculations. For example, in the calculations of Ref.~\cite{Bird:23},  only the ground rotational state was included in the basis, which amounts to neglecting the rotational structure of the molecule and the crucial anisotropy of the atom-molecule interaction at short range. Because of these drastic approximations, the previous calculations have only been able to formulate qualitative hypotheses (but not quantitative predictions) regarding the properties of atom-molecule MFRs.
By contrast, our calculations treat the atom-molecule interaction  anisotropy and the hyperfine structure of the collision partners exactly using converged rotational basis sets. This not only enables testing  the hypotheses formulated previously, but also makes  it possible to predict the properties of MFRs in atom-molecule systems with realistic interactions.
Our calculations also predict a new flavor of MFR due to the intramolecular spin-rotation interaction [see Fig.~\ref{fig:ch}]. 
}

To explore the short-range vs. long-range character of atom-molecule MFRs, we performed additional CC calculations with the short-range part of the Rb-SrF PESs scaled by a factor of  $\lambda=0.5$ (see \cite{SM} for details). 
Because the long-range parts of the scaled and unscaled PESs are the same, the scaling only changes the density of short-range complex states. As a result, it should not modify the density of MFRs if,  as proposed in Ref.~\cite{Frye:23}, the MFRs are due to the long-range states. 
If, on the other hand, the MFRs are due to the short-range states, then the PES scaling is expected to lead to a substantial decrease in MFR density according to the approximate scaling $\rho\simeq (\lambda)^{3/2}$  \cite{Frye:21}.
As shown in \cref{fig:scaled},
the number of MFRs decreases by 40\%  upon shallowing the PESs,  suggesting that MFRs in Rb~+~SrF collisions are predominantly caused by short-range complex states. This observation does not rule out the possibility that {some} MFRs could be caused by long-range states, which can have much narrower widths than those caused by the short-range complex states \cite{SM}, warranting further calculations and experimental measurements on near-threshold bound states of Rb-SrF, whose zero-energy crossings with collision thresholds give rise to MFRs \cite{Chin:10}. 
Such  calculations could be enabled by combining the efficient TRAM basis used here \cite{Tscherbul:23tram} with established techniques for solving the bound-state Schr\"odinger equation  \cite{Hutson:19b}. 


In summary, we have presented the first numerically exact CC calculations of MFRs in ultracold, strongly anisotropic Rb~+~SrF collisions in the rigid-rotor approximation. These calculations are free from the basis set truncation errors, which affected all previous calculations on such collisions, and enable us to obtain novel insights into the properties of atom-molecule MFRs.
First, we show that the MFRs can be classified into two types: (i) $J_r$-conserving broad resonances due to the spin-exchange splitting between the singlet and triplet PESs at short range,  and (ii) narrow resonances originating from $J_r$-breaking intramolecular spin-rotation interactions. While the former are similar to those which occur in alkali-metal atom collisions \cite{Chin:10}, the latter are unique to atom-molecule collisions.
 Second, we find that the density of MFRs in Rb~+~SrF collisions is comparable to that in collisions of ultracold alkali-metal atoms. We explain this surprising result in terms of the density of the adiabatic  states of the atom-molecule collision complex, which does not increase substantially as new rotational states are added to the CC basis.  The highly excited rotational states merely affect the shape of these adiabatic 
potentials rather than changing their density, thereby leaving the density of MFRs unaffected. We also find that
 the MFRs are predominantly due to the short-range complex states, supporting the hypothesis of Refs.~\cite{Mayle:12,Mayle:13}, although the existence of long-range complex states \cite{Frye:23} cannot be completely ruled out. 
 Our results suggest excellent prospects for numerically exact quantum scattering calculations of MFRs in ultracold highly anisotropic atom-molecule collisions, such as those probed in recent experiments \cite{Son:22,Park:23b} paving the way for future exploration of this exciting frontier of ultracold molecular physics.
 

 We thank Jeremy Hutson for a valuable discussion.
 This work was supported by the NSF CAREER program (Grant No. PHY-2045681) and by  the
U.S. Air Force Office of Scientific Research (AFOSR) under Contract No. FA9550-22-1-0361.
MK was supported by the National Science Centre, Poland (grant no. 2020/36/C/ST4/00508).
Computations were performed on the Niagara supercomputer at the
SciNet HPC Consortium located at the University of Toronto and at the Wroclaw Centre for Networking and Supercomputing (Grant No. 218).

%

\end{document}



\title{ \bf{Supplemental Material for \\  ``Magnetic Feshbach resonances in ultracold atom-molecule collisions''}}

\author{ Masato Morita$^{1}$, Maciej B. Kosicki$^{2}$, Piotr S. {\.Z}uchowski$^{3}$, Paul Brumer$^{1}$ and Timur V. Tscherbul$^{4}$ \\ }

\vspace{0.5cm}
\affiliation{ \vspace{0.3cm}
$^{1}$Chemical Physics Theory Group, Department of Chemistry, and Center for Quantum Information and Quantum Control, University of Toronto, Toronto, Ontario, M5S 3H6, Canada\\
$^{2}$Faculty of Physics, Kazimierz Wielki University, al. Powsta\'nc\'ow Wielkopolskich 2, 85-090 Bydgoszcz, Poland\\
$^{3}$Institute of Physics, Astronomy and Informatics, Nicolaus Copernicus University, Toru{\'n}, Poland,  87-100\\
$^{4}$Department of Physics, University of Nevada, Reno, Nevada, 89557, USA
}

\maketitle

\tableofcontents

\vspace{0.8cm}
In this Supplemental Material (SM), we provide technical details of our {\it ab initio}  and quantum scattering (coupled-channel, CC) calculations  on ultracold Rb~+~SrF collisions. 
%
\Cref{sec:SM_PESs} describes electronic structure calculations of the singlet and triplet interaction PES for Rb-SrF.
In \cref{sec:SM_Computation}, we provide computational details of our CC calculations, including the values of the parameters required to specify the Hamiltonian of the  Rb-SrF collision complex. 
To cross-validate our computational methodology and code based on the recently developed total rotational angular momentum (TRAM) basis \cite{Tscherbul_23}, we perform additional benchmark calculations using other established codes and compare the results in \cref{sec:SM_Program}. 

Numerical convergence of scattering observables with respect to  the number of rotational states and partial waves included in the CC basis set  is examined in \cref{sec:SM_Convergence},
 showing the predominance of the $l=0$ partial wave contribution at $E_\text{c}=0.1$~$\mu$K.
Additional details regarding the scaling of the short-range Rb-SrF PES and its effect on the density of magnetic Feshbach resonances (MFRs) are provided in \cref{sec:SM_scale}.
In \cref{sec:SM_initial}, we present more information on MFR spectra calculated  for the different initial hyperfine-Zeeman states of Rb and SrF. 
Finally, the adiabatic potential energy curves of the Rb-SrF collision complex are presented in \cref{sec:SM_adiabats}.

\setcounter{equation}{0}
\setcounter{figure}{0}

\vspace{0.5cm}
\section{\label{sec:SM_PESs} {\it Ab initio} calculations of {Rb-SrF} potentials}


Interactions between Rb and SrF can lead to either a low-spin singlet ($S=0$) or a high-spin triplet ($S=1$) PES. In collisions involving spin-polarized species, there is no spin exchange, confining the dynamics solely to the high-spin triplet state. In our previous work \cite{Morita_18} we focused on ultracold Rb~+~SrF collision dynamics on the triplet state, which is notably anisotropic due to the high polarity of SrF and the strong polarizability of Rb, leading to significant dipole-induced dipole interactions. The triplet PES was obtained from single-reference quantum chemistry calculations utilizing the ``gold-standard'' open-shell coupled cluster method with single, double, and noniterative triple excitations [CCSD(T)] and extensive basis sets \cite{Kosicki:17}. 
However, accurately modeling the global PES for the low-spin singlet state necessitates a multi-reference approach, and remains a substantial challenge. 
The methods like multireference configuration interaction or multi-reference perturbation theory often fail to capture the asymptotics of the interaction accurately due to size-consistency issues. Nonetheless, we found that using complete active space self-consistent field (CASSCF) calculations, where the active space consists of  bonding orbitals paired with their antibonding counterparts, allows for the accurate computation of the low-spin PES. 

Our strategy involves augmenting the high-quality triplet state obtained via CCSD(T) with a differential potential calculated at the CASSCF level. We performed these calculations using the augmented core-valence correlation-consistent basis set (aug-cc-pCVQZ) for the F atom, and tailored small-core relativistic energy-consistent pseudopotentials (ECP28MDF) for Rb and Sr, complemented by even-tempered functions for each angular momentum and truncated to spdf orbitals~\cite{Lim:2005, Lim:2006}. We examined the potential energy surfaces for Rb($^2S$) + SrF($^2\Sigma^+$) across both low-spin ($^1$A') and high-spin ($^3$A') electronic ground states using the Jacobi coordinates $R$ and $\theta$ (see the main text).

Assuming that the SrF monomer is rigid, we computed the energy of both lowest electronic states in each spin state as a function of the COM-atom distance  $R$ and the $\theta$ angle at a constant SrF bond length (2.075 \AA), corresponding to the experimentally determined equilibrium geometry \cite{Herzberg:79}. The radial and angular grids ranged from $R=2$ to 10~\AA~ and $\theta=0-180^\circ$ in steps of $\theta=5^\circ$ and 0.25~\AA, respectively. With such an approach the  low-spin surface properly recovers the long-range behavior for both spin states, which should have the identical inverse-power expansion, including short-range dispersion and polarization interactions. Note that a similar approach was used before for the low-spin PESs of O$_2$-O$_2$  and NH-NH \cite{Janssen2009,Bartolomei2008,Zuchowski2008}.

\begin{figure}
    \centering
\subfloat[] {
  \includegraphics[width=0.44\linewidth]{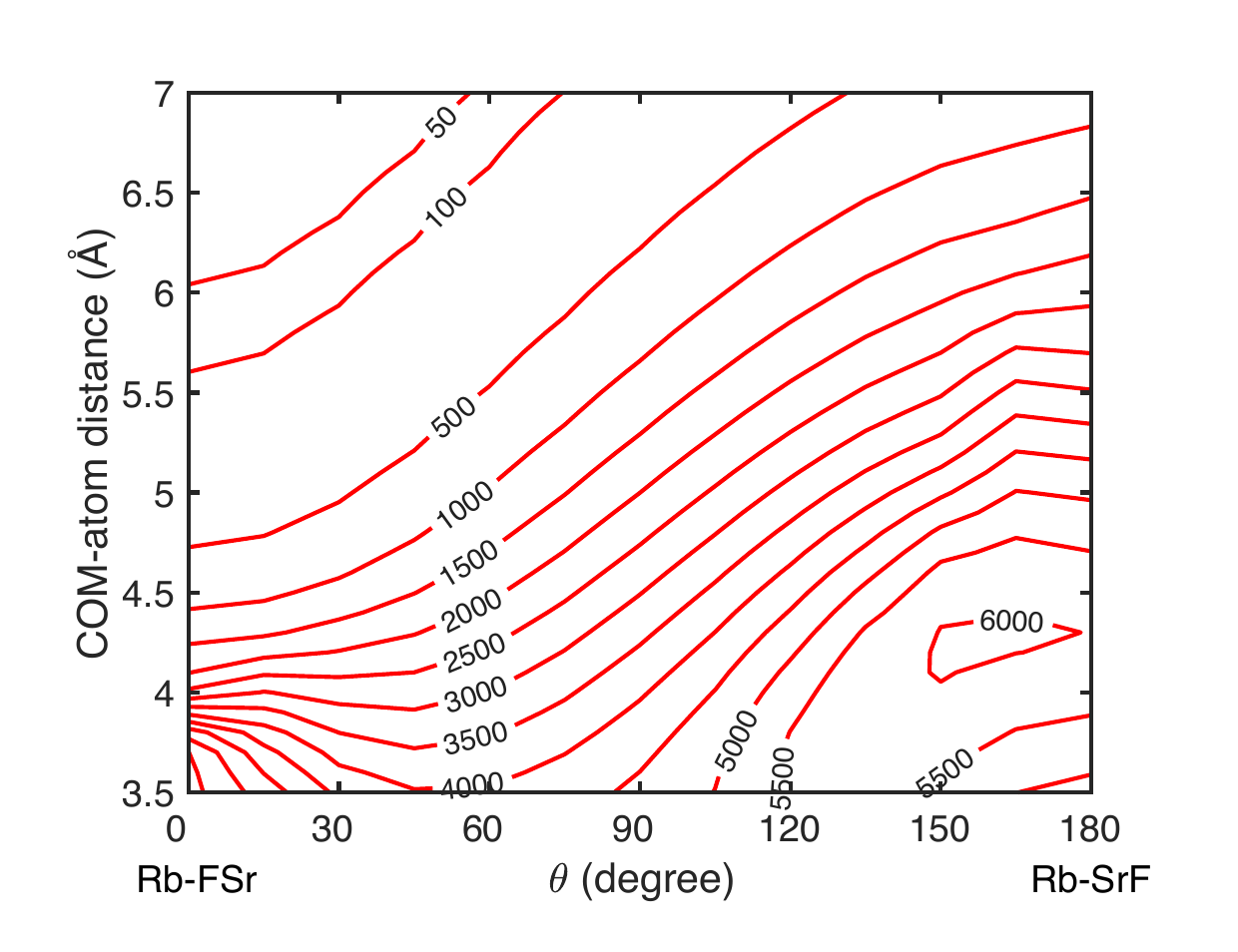}
}
\subfloat[]{
  \includegraphics[width=0.49\linewidth]{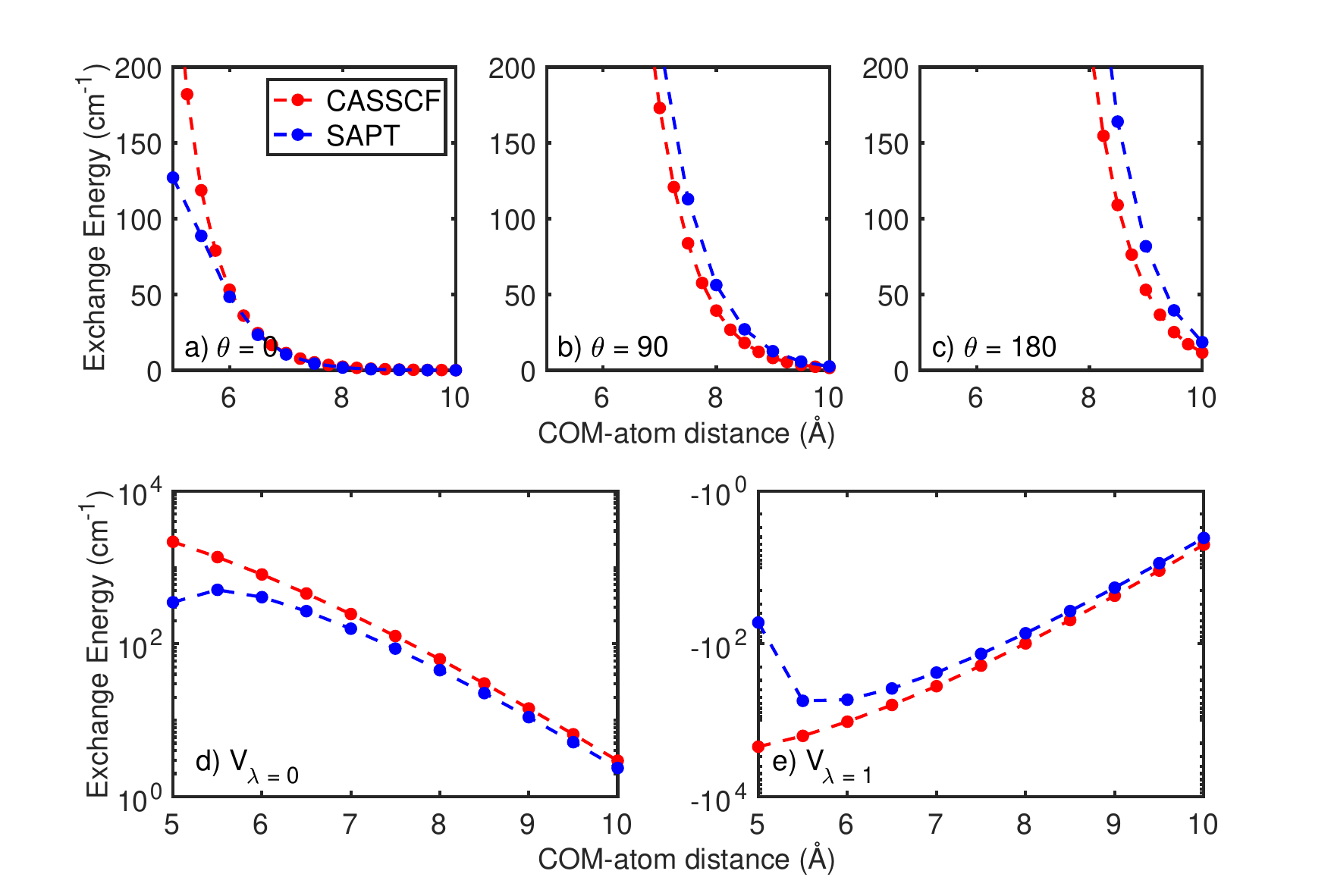}
}
 \caption{(a) Contour plot of the difference PES for Rb-SrF as a function of $R$ and $\theta$. (b) Difference PES plotted as a function of $R$ for selected values of $\theta$ [top panels] and the Legendre moments $V_0(R)$ and $V_1(R)$ [bottom panels]. Red circles -- CASSCF calculations, blue circles -- SAPT calculations. }
 \label{fig:PES}
\end{figure}

 A two-dimensional plot of the difference PES $V_\text{diff}(R,\theta)=V^{S=1}(R,\theta)-V^{S=0}(R,\theta)$ is shown in \cref{fig:PES}(a). The difference PES  reaches values as high as 6000 cm$^{-1}$, and  exhibits a tremendous angular anisotropy. The difference PES becomes particularly large (i.e., the singlet PES becomes particularly deep)  when the Rb atom approaches the  Sr end of the SrF molecule. This is due to the fact that, within the SrF molecule, the $5s$ electron of the Sr atom is captured by the fluorine atom, forming a closed-shell F$^-$ anion, and the Sr$^+$ ion exhibits a large spin density. In \cref{fig:PES}(b), we show the radial profiles of the difference PES for 0, 90 and 180 degrees along with the Legendre moments
  $$V_\lambda(R)=\frac{1}{2}\left(2\lambda+1\right)\int_{-1}^{1} V_\text{diff}(R,\theta) P_\lambda(\cos(\theta)) d\cos \theta,$$ 
  which characterize its anisotropy.  In particular,  $V_0(R)$  corresponds to the isotropic part of the difference PES and  $V_1(R)$  to the leading anisotropic term.
 
Recently, one of us developed new theory for calculations of the first-order interaction energies for general spin quantum numbers, \cite{Patkowski2018} in the framework of the symmetry-adapted perturbation theory (dubbed as spin-flip SAPT). We performed the SF-SAPT calculations to assess the quality of the  CASSCF approach. The agreement between the difference PESs calculated using SF-SAPT and CASSCF is very good for $R > 6$~\AA. The discrepancy between these two approaches at short $R$ is expected and results from inadequate description of the short-range interaction by the SF-SAPT theory.
The  discrepancy  can have a major effect on the positions and widths of the individual MFRs (which are well known to be highly sensitive to tiny variations of the interaction PES \cite{Bodo:04,Hutson:07,Suleimanov:16,Morita_19,Morita:23}. However, we expect the overall MFR pattern and the MFR density to be adequately described by these PES, since they are both realistically deep and anisotropic, and display the correct long-range behavior.


\vspace{0.5cm}
\section{\label{sec:SM_Computation} Coupled-channel methodology: Computational details}

The  quantum scattering methodology used in this work 
has been presented in Ref.~\cite{Tscherbul_23}, so we only outline the most essential details here.
The Hamiltonian of the Rb($^2\mathrm{S}$) - SrF($^{2}\Sigma$)  collision complex may be written as (in atomic units)
\begin{equation}\label{eq:H}
\hat{H}=-\frac{1}{2\mu R}\frac{\partial^2}{\partial R^2}R + \frac{\hat{L}^2}{2\mu R^2}
+ \hat{V}_\text{int} 
+ \hat{H}_\text{mol} + \hat{H}_\text{at}.
\end{equation}
where the $^{87}$Rb - $^{88}$Sr$^{19}$F reduced mass $\mu = 47.93760469$ amu. 
The Hamiltonian $\hat{H}_\text{mol}$ of the  $^{88}$Sr$^{19}$F molecule is 
\begin{equation}\label{eq:Hmol}
\hat{H}_\text{mol}= \hat{H}_\text{mol}^\text{rot} + \hat{H}_\text{mol}^\text{spin} + \hat{H}_\text{mol}^\text{spin-rot},
\end{equation}
where $\hat{H}_\text{mol}^\text{rot}  = B_e\hat{N}^2$ is the rigid rotor Hamiltonian and $B_e=0.24975935$~cm$^{-1}$ is the rotational constant  \cite{Sheridan_09}. The molecular spin Hamiltonian $\hat{H}_\text{mol}^\text{spin}  = g_S\mu_0 B \hat{S}_Z +  a\hat{\mathbf{I}}\cdot\hat{\mathbf{S}}$ accounts for the interaction of electron spin with an external magnetic field $B$ directed along a space-fixed $z$-axis, and for the isotropic hyperfine interaction between the  electron and nuclear spins of SrF described by the operators $\hat{\mathbf{S}}$ and $\hat{\mathbf{I}}$ ($a=0.0035748333$ cm$^{-1}$  \cite{Childs_81}). 
The third term in \cref{eq:Hmol} describes the spin-rotation interactions $\hat{H}_\text{mol}^\text{spin-rot}= \gamma_\text{sr} \hat{\mathbf{N}}\cdot \hat{\mathbf{S}} + \frac{c\sqrt{6}}{3}\bigl{(}\frac{4\pi}{5}\bigr{)}^{1/2}  \sum_{q} (-1)^qY_{2-q}(\theta_r,\phi_r) [\hat{\mathbf{I}}\otimes \hat{\mathbf{S}}]^{(2)}_q$, which include the electron spin-rotation and anisotropic hyperfine interactions, parametrized by the coupling constants $\gamma_\text{sr}= 0.0024974$ cm$^{-1}$ \cite{Sheridan_09} 
and $c=0.0010096$ cm$^{-1}$ \cite{Childs_81} respectively. Finally, the internal structure of atomic Rb($^2$S)  is described by the Hamiltonian $\hat{H}_\text{at}  = g_S\mu_0 B \hat{S_a}_{z} + A_a\hat{\mathbf{I}}_a\cdot\hat{\mathbf{S}}_a$, where ${\mathbf{S}}_a$ and $\hat{\mathbf{I}}_a$ are the atomic electron and nuclear spin operators, and $A_a=0.11399$ cm$^{-1}$ is the hyperfine constant \cite{Arimondo_77}.

The CC equations formulated in the TRAM basis \cite{Simoni:06,Tscherbul_23} are solved numerically using the log-derivative propagation method \cite{Johnson_73,Manolopoulos_86}. 
The propagation is carried out from $R_\text{min}=5.0\ a_0$ to $R_\text{max}=800.0\ a_0$. This range of $R$ is divided into 4 intervals, with different propagation steps $\Delta R$ used within each interval as follows: $\Delta R=0.002 \ a_0$ in  $5\le R \le 13\ a_0$, $\Delta R=0.005 \ a_0$ in $13 \le R\le 17\ a_0$, $\Delta R=0.01 \ a_0$ in $17\le R \le20\ a_0$, and  $\Delta R=0.1 \ a_0$ in 20 $\le R \le 800\ a_0$.\\

\vspace{0.5cm}
\section{\label{sec:SM_Program} Code verification}

An extended version of the TRAM code developed in Ref.~\cite{Tscherbul_23} for He~+~YbF collisions is used to perform the CC calculations reported in this work. We introduced the following modifications  to  adapt the code to the much more challenging case of Rb~+~SrF collisions. First, we added the magnetic dipole-dipole interaction to the potential coupling matrix following Eq.~(28) of Ref.~\cite{Tscherbul_23}.
Second, we implemented two adiabatic Rb-SrF PESs to calculate the matrix elements of the interaction potential in the TRAMP basis as described in Sec. IIB of  Ref.~\cite{Tscherbul_23}. Third, we incorporated the hyperfine structure of the Rb atom into the basis.

To validate our newly developed atom-molecule scattering code, we performed test calculations of inelastic (spin relaxation) cross sections  from the fully spin-stretched initial states  of Rb and SrF.
\Cref{table:program} shows that these calculations are in excellent agreement with those of independent benchmark calculations using computer codes, which employ different basis sets \cite{Krems_04,Tscherbul_10, Morita_18}. In these tests, we omitted the hyperfine structure of both the collision partners since the previous codes did not take it in into account.
Because the inelastic transitions from the fully spin-stretched initial states are mainly driven by the magnetic dipole-dipole interaction \cite{Morita_18}, the agreement provides a strong evidence for the correctness of our implementation of the magnetic dipole interaction. 

\begin{table}[t!]
\centering
\caption{Total inelastic cross section (in $\mathrm{\AA}^2$) for Rb($M_S=1$)+SrF($N=0$, $M_S=1$), $E_\text{c}=0.1 \, \mu$K, $N_\text{max}=3$, $M_\text{tot}=1$. The ``TRAM'', ``TAM'' and ``Uncoupled SF'' columns list the results obtained using the TRAM basis \cite{Tscherbul_23}, total angular momentum (TAM) basis  \cite{Tscherbul_10}, and the fully uncoupled basis  \cite{Krems_04,Volpi_02}. The labels SF and BF denote the space-fixed frame and body-fixed frame versions of the basis.}
\begin{tabular}[t]{>{\raggedright}m{0.2\linewidth}>{\raggedright\arraybackslash}m{0.2\linewidth}>{\raggedright\arraybackslash}m{0.2\linewidth}>{\raggedright\arraybackslash}m{0.2\linewidth}} 
\hline
Magnetic field  (G)          &      TRAM(SF)       &    TAM (BF)        &  Uncoupled (SF)     \\
\hline
1000&14.8186&14.8210&14.8193 \\
100&76.6500&76.6490&76.6279\\
10&381.417& 381.404&381.372\\
\hline
\label{table:program}
\end{tabular}
\end{table}

To test the hyperfine part of our TRAM code, we performed benchmark calculations using the fully  uncoupled basis set \cite{Volpi_02, Krems_04} augmented with the hyperfine basis states of Rb and SrF as described in  Ref.~ \cite{Tscherbul_07}. We used a limited uncoupled CC basis consisting of three lowest rotational states of SrF (CC calculations with $N_\text{max}\ge  15$  are computationally unfeasible in the uncoupled basis  \cite{Tscherbul:11}, so we only use it here for benchmarking purposes).

In \cref{table:program_hpf}, we list the inelastic cross sections for Rb~+~SrF calculated with the hyperfine structure of the collision partners  included for the energetically highest initial states of Rb and SrF [see Fig.~1(c) of the main text].
The excellent agreement between the TRAM and uncoupled basis results validates the implementation of the hyperfine interactions of both the collision partners  in our  TRAM code. 

\begin{table}[b!]
\centering
\caption{Total inelastic cross section (in units of $\mathrm{\AA}^2$) for Rb($|8\rangle$)+SrF($|4\rangle$) collisions at $E_\text{c}=10^{-6}$ cm$^{-1}$,   $N_\text{max}=2$, and the total angular momentum projection $M_\text{tot}=3$. The column labels are the same as in \cref{table:program}. }
\begin{tabular}[t]{>{\raggedright}m{0.2\linewidth}>{\raggedright\arraybackslash}m{0.2\linewidth}>{\raggedright\arraybackslash}m{0.2\linewidth}}
\hline
Magnetic field (G)           &      TRAM(SF)        &  Uncoupled (SF) \\
\hline
1000 &97.6986&97.9521\\
1      &1949.23&1950.43\\
\hline
\label{table:program_hpf}
\end{tabular}
\end{table}

\section{\label{sec:SM_Convergence} Basis set convergence}

\begin{figure}[t!]
\begin{center}
\includegraphics[height=0.29\textheight,keepaspectratio]{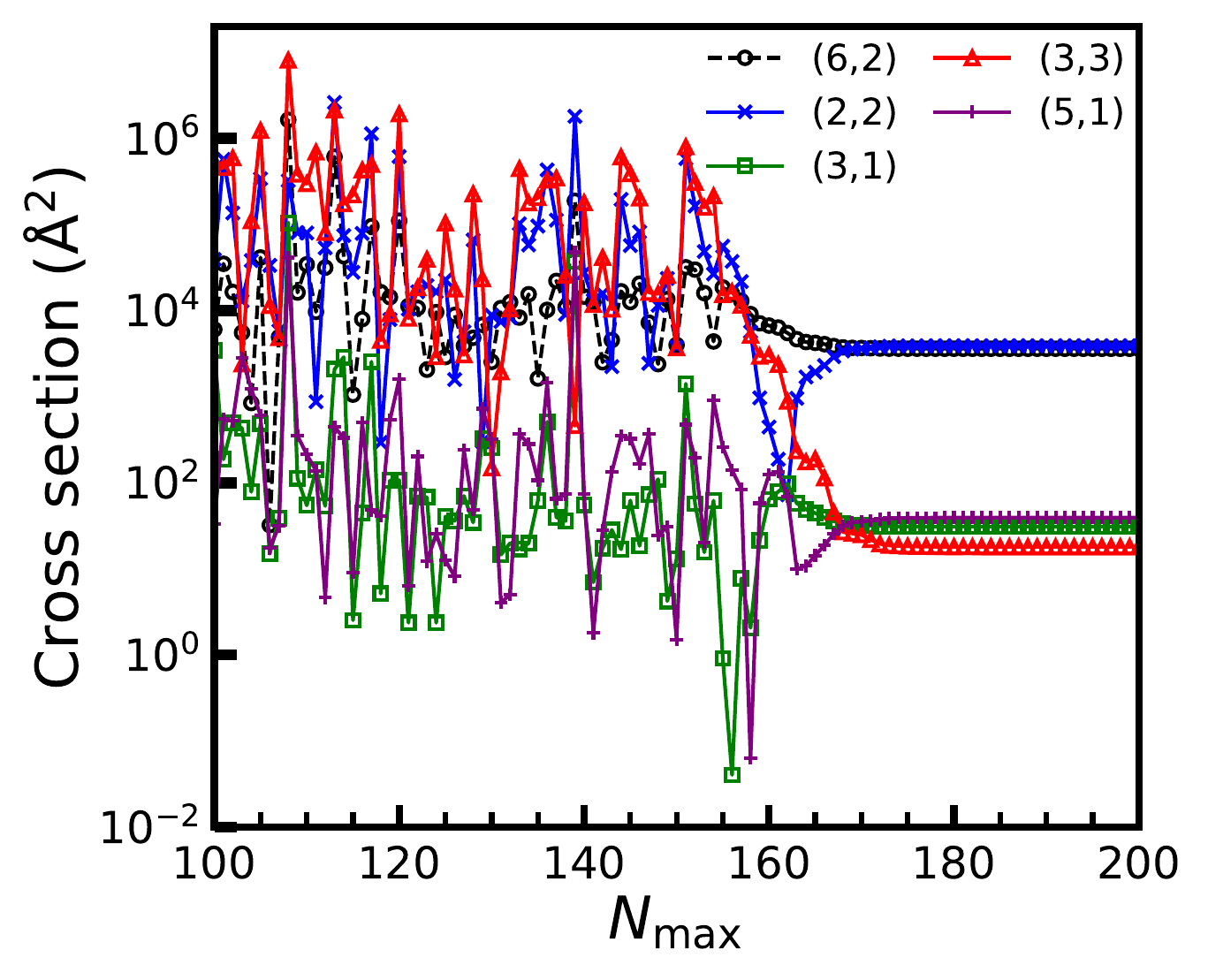}
\end{center}
\caption{
Convergence of the cross sections with respect to rotational basis set size ($N_\text{max}$) for   Rb($|6\rangle$)+SrF($|2\rangle$) collisions. 
The final states are indexed as ($i,j$) where $i$ and $j$ are the labels of the hyperfine-Zeeman states of Rb and SrF as specified in the main text. 
 The black dashed curve with open circles, (6,2), denotes the elastic cross section.
$E_\text{c} = 0.1\,\mu$K, $B = 1000$~G, $J_r^\text{max} = 0$, and $M_\text{tot}=-1$. }
\label{Fig_SM_Nconv}
\end{figure}

\begin{figure}[h!]
\begin{center}
\includegraphics[height=0.52\textheight,keepaspectratio]{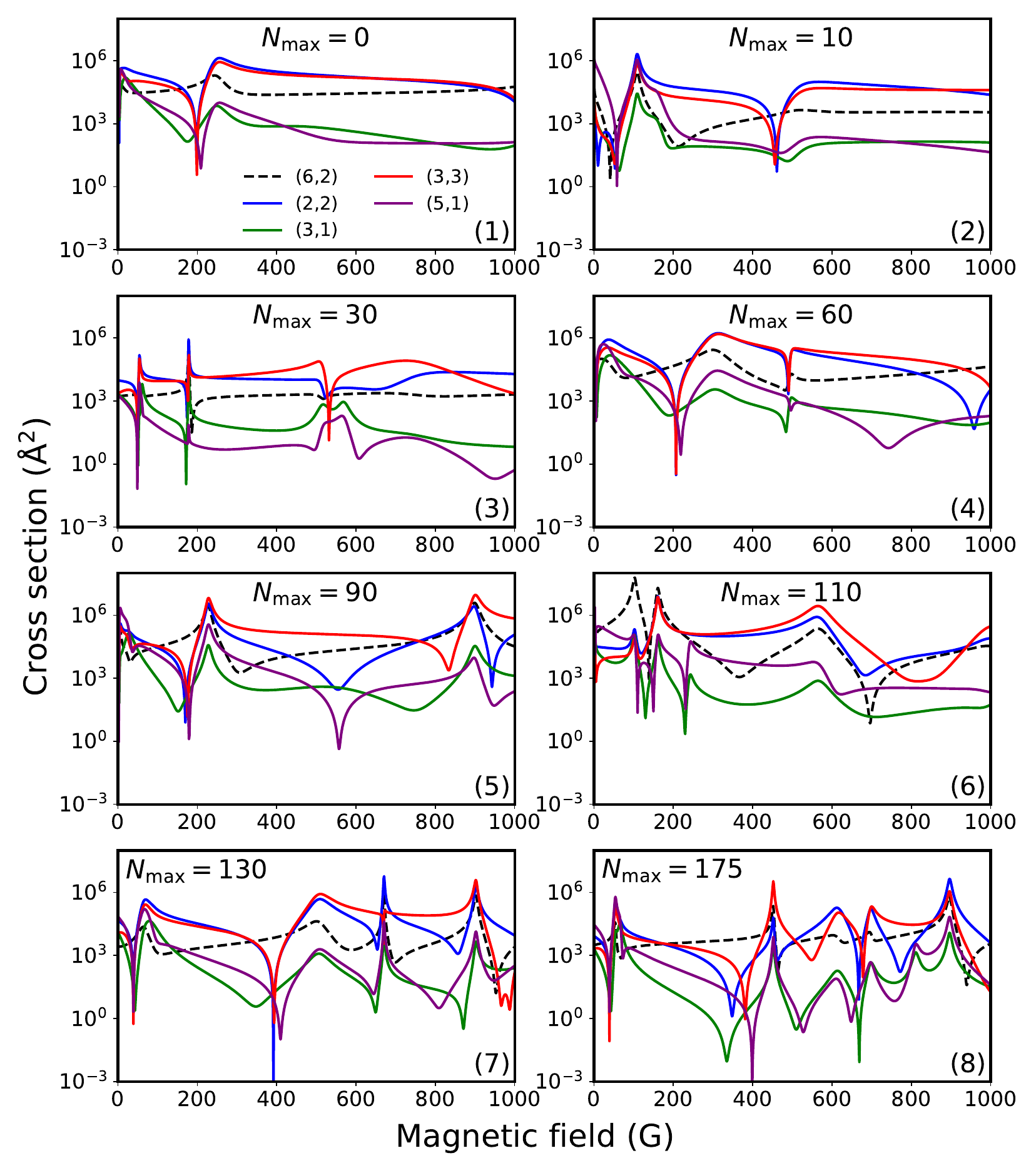}
\end{center}
\caption{
Rotational basis set ($N_\text{max}$) dependence of MFR spectra  for  Rb($|6\rangle$)+SrF($|2\rangle$) collisions.
The final states are indexed as ($i,j$) where $i$ and $j$ are the labels of the hyperfine-Zeeman states of Rb and SrF as specified in the main text. 
 The black dashed curve  denotes the elastic cross section.
$E_\text{c} = 0.1\,\mu$K, $J_r^\text{max} = 0$, and $M_\text{tot}=-1$.  }
\label{Fig_SM_NconvB}
\end{figure}

The Rb-SrF interaction is deep and highly anisotropic (see Sec.~I above), causing strong coupling between the different rotational basis states of SrF. As with other  highly anisotropic atom-molecule collisions such as Li~+~CaH and Li~+~SrOH explored before  \cite{Tscherbul:11,Morita:17,Morita_18} this makes it necessary to include a large number of excited rotational states in the TRAM basis to achieve numerical convergence. The basis set cutoff parameter  $N_\text{max}$ is determined by an interplay of the potential anisotropy, well depth and the rotational constant of the molecule.

In \cref{Fig_SM_Nconv}, we show the $N_\text{max}$ dependence of the cross sections for Rb($|6\rangle$)+SrF($|2\rangle$) collisions (see the main text for the  labeling of the internal states of Rb and SrF). 
Due to the deep well and high anisotropy of the singlet PES, one expects a large number of rotational states of SrF to be required for convergence, as is indeed observed in \cref{Fig_SM_Nconv}. Specifically, convergence is reached at $N_\text{max}=175$, which is much larger than in the previous calculations on the triplet PES  ($N_\text{max}=125$) \cite{Morita_18}.  
Interestingly, we observe significant variations in the cross sections near the convergence limit ($N_\text{max}=155-170$), demonstrating the importance of  basis set convergence for the quantitative prediction of scattering  observables. 

\vspace{0.5cm}

\Cref{Fig_SM_NconvB} shows the MFR spectra for Rb($|6\rangle$)+SrF($|2\rangle$) collisions calculated using rotational  basis sets of varying size. 
While the cross sections fluctuate strongly as a function of $N_\text{max}$,  the number of major resonances between $B=0$ and 2000~G increases only moderately (from 3 to 4) in going from $N_\text{max}= 30$ to the fully converged basis limit ($N_\text{max}= 175$).
As discussed in the main text, adding basis functions with higher rotational quantum numbers 
 causes slight shifts in the existing adiabatic potentials (see below), which affects the individual resonance positions but not their overall density.

\begin{figure}[h!]
\begin{center}
\includegraphics[height=0.25\textheight,keepaspectratio]{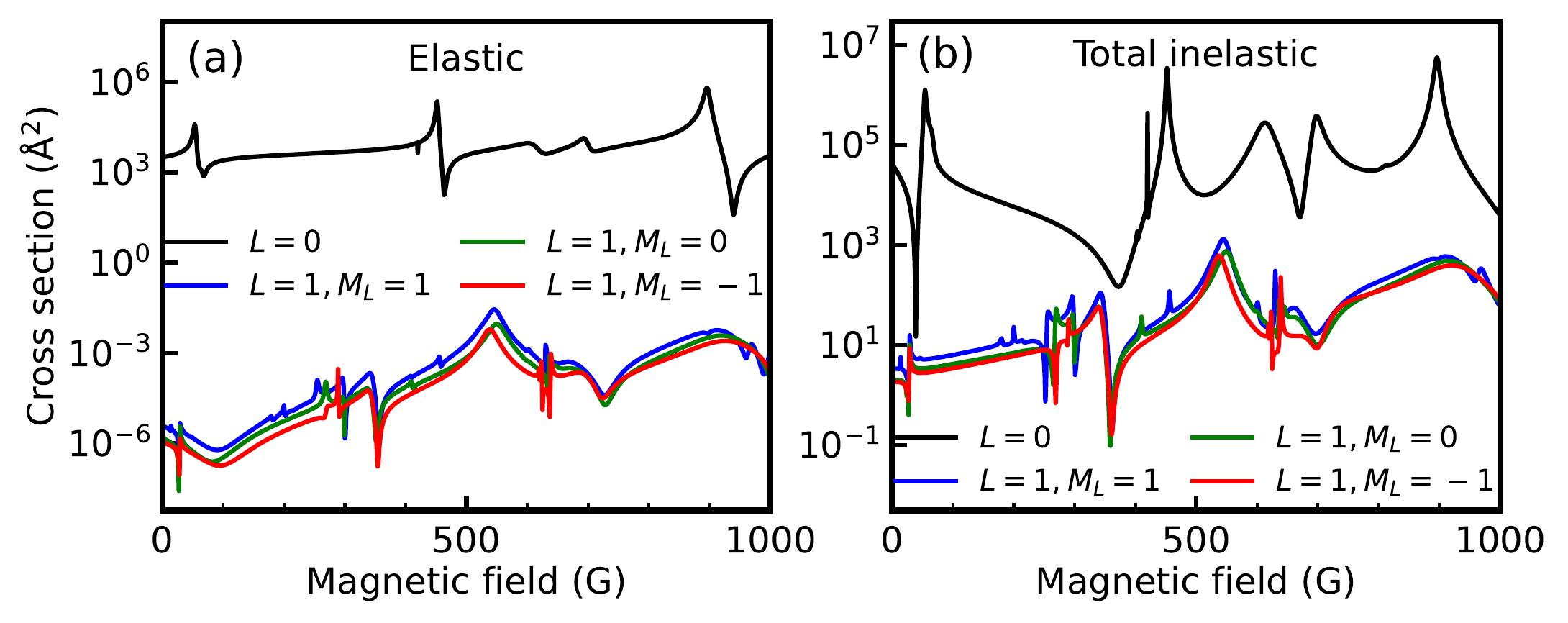}
\end{center}
\caption{
Partial wave decomposition of elastic and total inelastic cross sections for Rb($|6\rangle$)+SrF($|2\rangle$) collisions. 
$E_\text{c} = 0.1\, \mu$K,  $J_r^\text{max} = 1$, and $N_\text{max}=175$. }
\label{Fig_SM_partial}
\end{figure}

\vspace{0.5cm}
In Figs.~1(a) and (b) of the main text, we show the $J_r^\text{max}=0$ and  $J_r^\text{max}=1$ calculations at $E_\text{c}=0.1\, \mu$K, and focus on  $J_r^\text{max}=0$ calculations  in the remainder of the main text.  These latter calculations take into account only the incoming $s$-wave component of atom-molecule scattering.
To verify that higher partial waves make a negligible contribution at this collision energy,  we directly calculated the three $p$-wave components of the cross sections. As shown in \cref{Fig_SM_partial},  both the elastic and inelastic cross sections are dominated by the $s$-wave component throughout the entire  range of  magnetic fields.

 
\vspace{0.8cm}
\section{\label{sec:SM_NoSE} Effect of the spin-exchange interaction}

\begin{figure}[h!]
\begin{center}
\includegraphics[height=0.2\textheight,keepaspectratio]{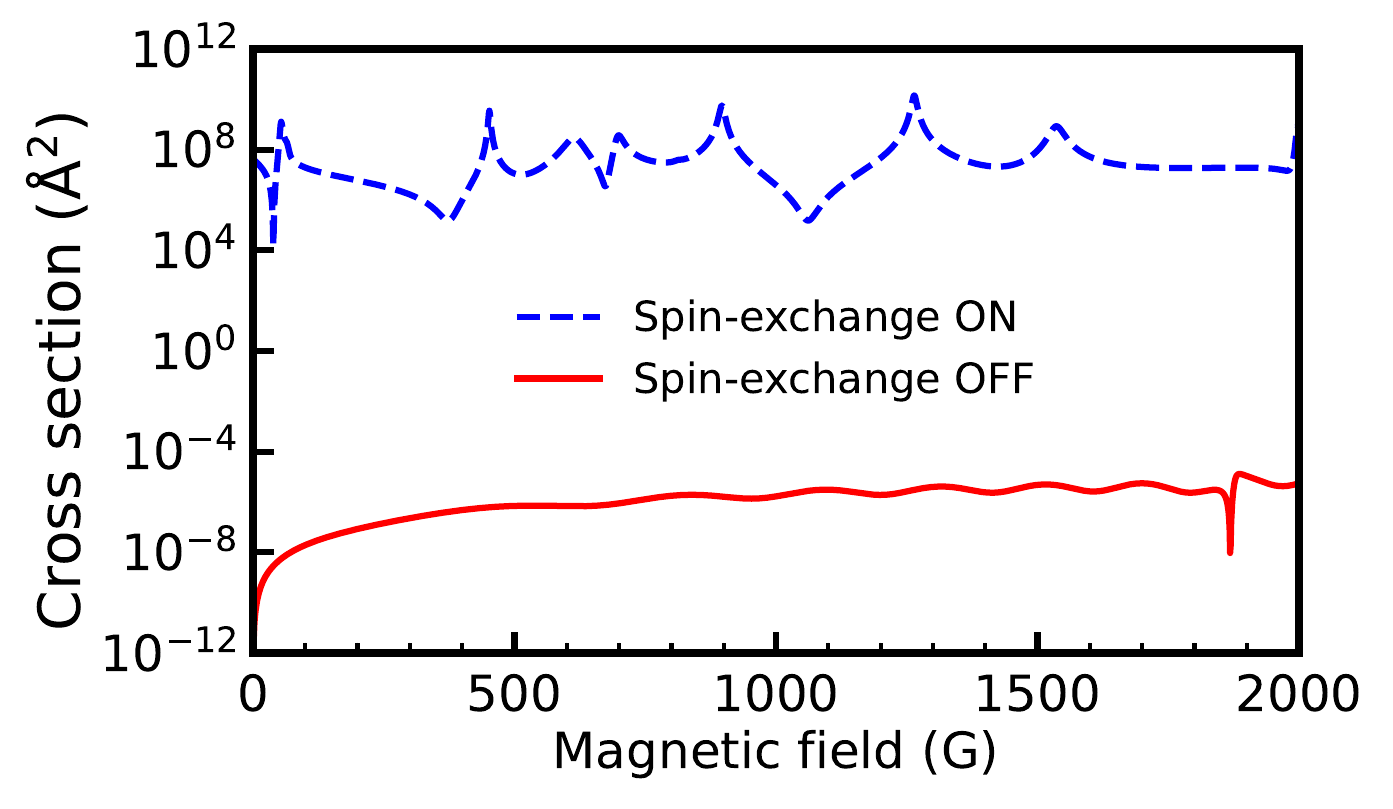}
\end{center}
\caption{
Total inelastic cross sections for Rb($|6\rangle$)+SrF($|2\rangle$) collisions calculated with (blue dashed curve) and without (red solid curve) the spin-exchange interaction. $E_\text{c} = 0.1\, \mu$K, $J_r^\text{max} = 0$, and $N_\text{max}=175$. }
\label{Fig_SM_se}
\end{figure}

\begin{figure}[b!]
\begin{center}
\includegraphics[height=0.35\textheight,keepaspectratio]{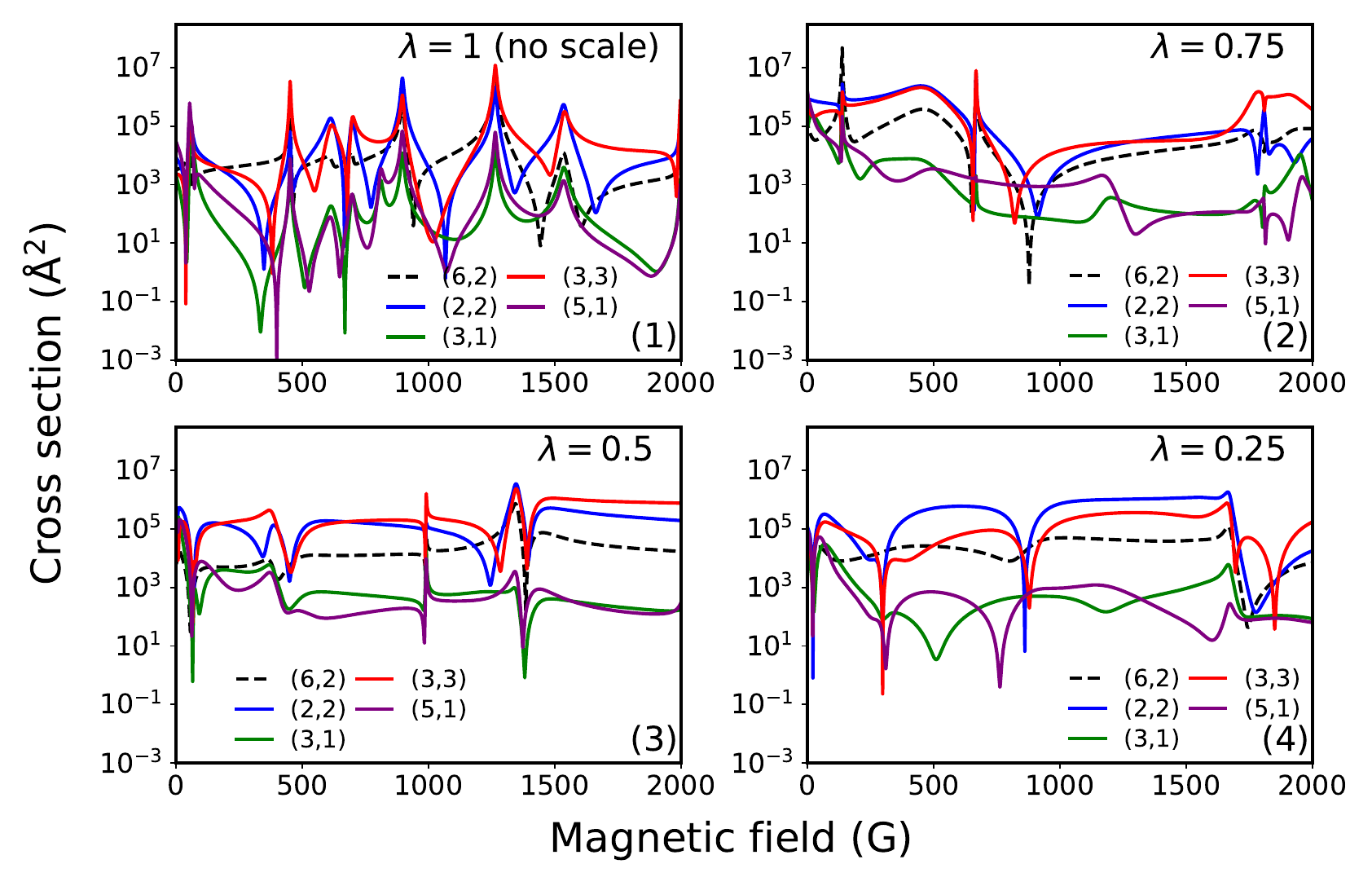}
\end{center}
\caption{
MFR spectra for ultracold Rb($|6\rangle$)+SrF($|2\rangle$) collisions calculated with unscaled $(\lambda=1)$ and scaled $(\lambda\ne 1)$  short-range PESs. Both singlet and triplet PES  are scaled by multiplying their short-range parts by a constant scaling factor $\lambda$.  The elastic cross section is shown as the black dashed line.
$E_\text{c} = 0.1\, \mu$K, $J_r^\text{max} = 0$, and $N_\text{max}=175$. 
 }
\label{Fig_SM_scale}
\end{figure}
Previous quantum scattering calculations on ultracold atom-molecule collisions in an external magnetic field have focused on  spin-stretched  initial states (see, e.g., Refs.~\cite{Morita_17,Morita_18,Karman:23,Park:23b})
Such calculations can be performed on a single PES with the highest spin multiplicity.
To our knowledge, this work is the first rigorous CC study of ultracold atom-molecule collisions involving non fully-stretched initial states.

The spin-exchange interaction, given by the difference between the singlet and triplet PESs, plays a crucial role in the formation of MFRs  in ultracold collisions of non-fully stretched initial states  of alkali-metal atomss \cite{Chin:10}.
To elucidate the role of the spin-exchange interaction in  ultracold Rb+SrF collisions, we performed test CC calculations with  the singlet PES $V_{S=0}(R,\theta)$ made identical to the triplet PES $V_{S=1}(R,\theta)$, effectively turning off the spin-exchnage  interaction. 
As shown in \cref{Fig_SM_se}, the total inelastic cross section for Rb($|6\rangle$)+SrF($|2\rangle$) collisions is dramatically suppressed when the  spin-exchange interaction is turned off, demonstrating the crucial role of this interaction in inducing inelastic scattering. Furthermore, neglecting this interaction alters the selection rules for inelastic transitions, making the $(2,2)$ final channel dominant and all other transitions vanishing identically.

{To estimate the effect of the long-range magnetic dipole-dipole interaction  on MFR spectra, we  performed additional calculations with this interaction omitted. The calculations show that
the magnetic dipole-dipole interaction  does not substantially alter the cross sections plotted in Figs. 1-3 of the main text. For example, the relative difference between the cross sections calculated with and without the magnetic dipole-dipole interaction with $J_r^\text{max}=0$ is less than 0.0017\% for all the initial states shown in Fig.~2 of the main text. Therefore, ultracold Rb~+~SrF scattering can be described by a single s-wave scattering length. }


\begin{figure}[b!]
\begin{center}
\includegraphics[height=0.6\textheight,keepaspectratio]{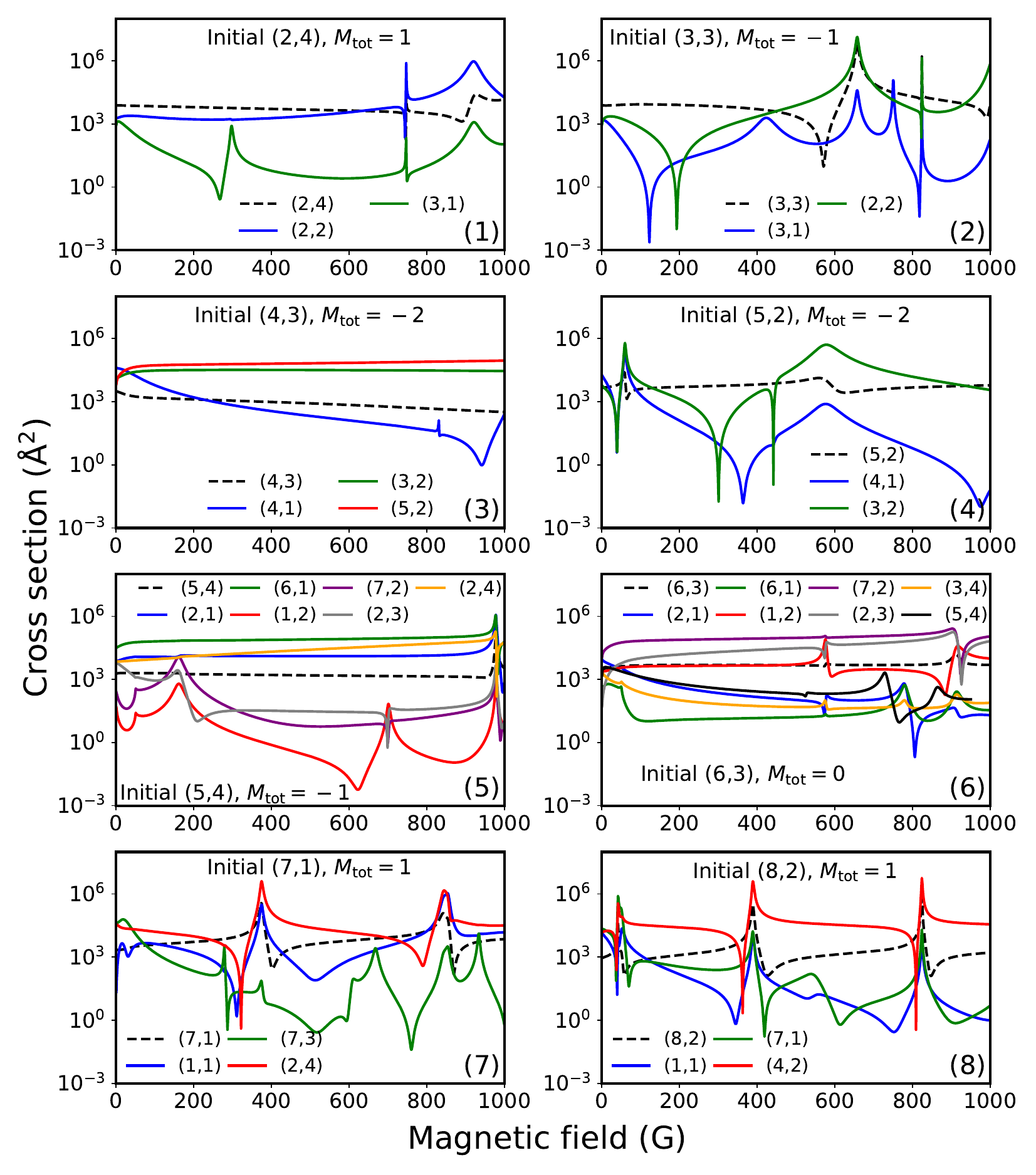}
\end{center}
\caption{
 MFR spectra for ultracold Rb~+~SrF collisions calculated for different initial states.
The black dashed curve with open circles denotes the elastic cross section.
$E_\text{c} = 0.1 \mu$K, $J_r^\text{max} = 0$, and $N_\text{max}=175$. }
\label{Fig_SM_initial}
\end{figure}

\section{\label{sec:SM_scale} Resonance density and potential scaling}

\Cref{Fig_SM_scale}  shows MFR spectra for Rb-SrF collisions computed with  the short-range Rb-SrF interaction PES  multiplied  by a constant scaling factor $\lambda$. 
We observe a gradual decrease in the resonance density as the value of $\lambda$ is decreased from 1 to 0.25, suggesting that the major resonance peaks are due to the short-range states localized in the potential well.

\section{\label{sec:SM_initial} Different initial states}

\Cref{Fig_SM_initial}  shows MFR spectra calculated for the different initial states of Rb and SrF. 
While the cross sections exhibit significant initial and final state dependence, the density of resonance is not sensitive to the initial and final states.

\section{\label{sec:SM_adiabats} Adiabatic potentials}

\begin{figure}[h!]
\begin{center}
\includegraphics[height=0.45\textheight,keepaspectratio]{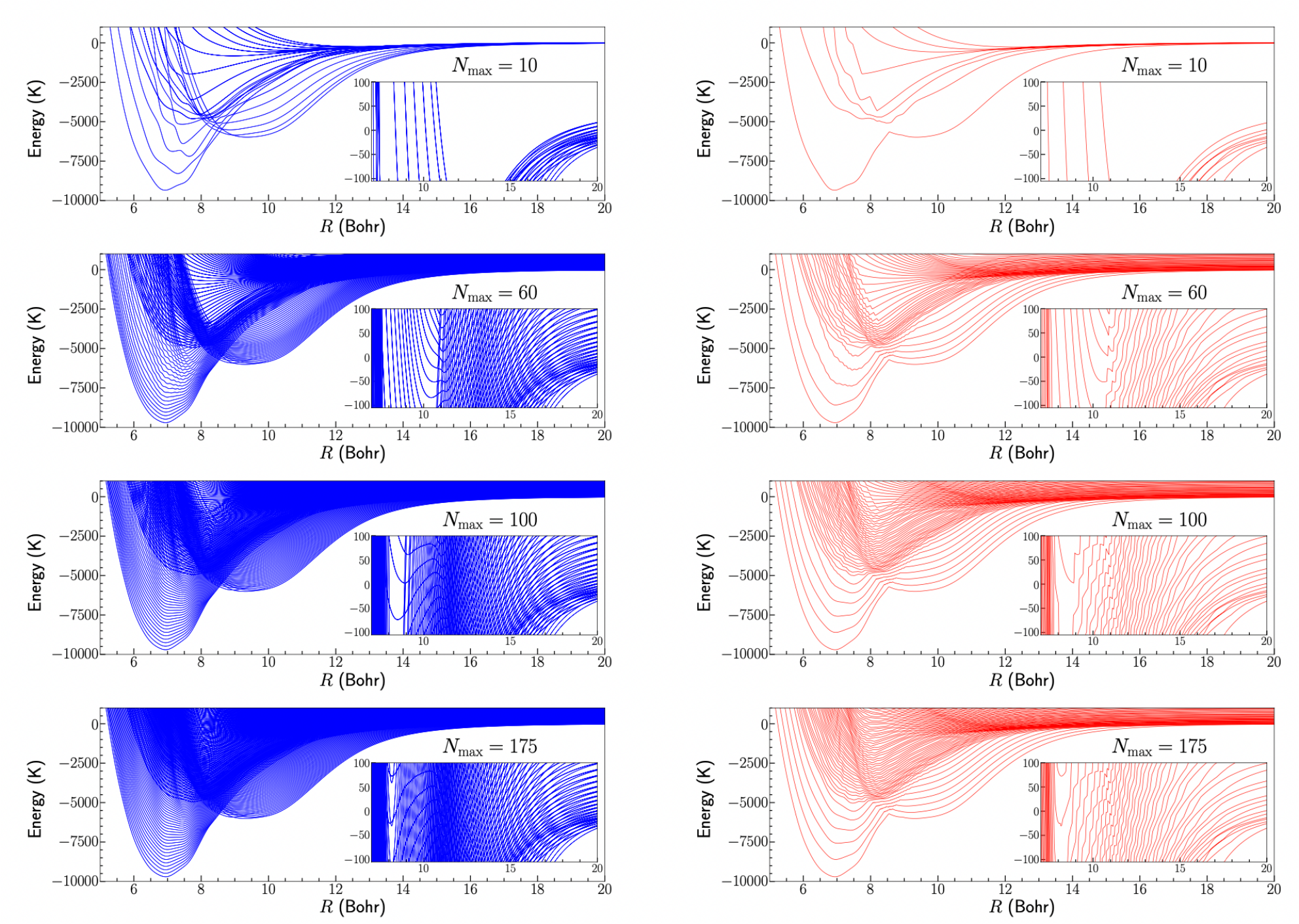}
\end{center}
\caption{
(Left panels) Adiabatic potentials for Rb-SrF calculated for $B = 1000$~G, $J_r^\text{max} = 0$, $M_\text{tot} = -1$, and $N_\text{max}=175$.
Every 10-th adiabatic potential is plotted in the right panels.  }
\label{Fig_SM_adiabats}
\end{figure}

 \Cref{Fig_SM_adiabats}  displays the adiabatic potential energy curves obtained by diagonalizing the Rb-SrF Hamiltonian with the radial kinetic energy term omitted.
We observe that the adiabatic potentials correlating with the $N=0$ threshold  (relevant to the MFRs studied in this work) converge quickly with $N_\text{max}$ except in the vicinity of avoided crossings at short range. 
Increasing $N_\text{max}$ beyond $\simeq 60$ does not introduce any new  features (such as avoided crossings) in the vicinity of the $N=0$ threshold. As such, even though highly rotationally excited states can modify the shape of the existing adiabatic potentials near threshold, they do not affect the density of closed-channel bound states, which give rise to MFRs.

%